%% file: main.tex
\documentclass{article}

\PassOptionsToPackage{numbers, compress}{natbib}

    \usepackage[final]{tackling_climate_workshop_style}

\usepackage[utf8]{inputenc} 
\usepackage[T1]{fontenc}    
\usepackage{hyperref}       
\usepackage{url}            
\usepackage{booktabs}       
\usepackage{amsfonts}       
\usepackage{nicefrac}       
\usepackage{microtype}      
\usepackage{graphicx}       
\usepackage{siunitx}        
\usepackage{makecell}
\usepackage{multirow}
\usepackage{wrapfig}

\input{math_commands.tex}

\title{Understanding Ice Crystal Habit Diversity with Self-Supervised Learning}

\author{%
  Joseph Ko \\
  Columbia University \\
  New York, New York \\
  \texttt{jk473@columbia.edu} \\
  \And
  Hariprasath Govindarajan  \\
  Qualcomm Auto Ltd Sweden Filial \& \\
  Linköping University \\
  Linköping, Sweden \\
  \texttt{hargov@qti.qualcomm.com} \\
  \AND
  Fredrik Lindsten \\
  Linköping University \\
  Linköping, Sweden \\
  \texttt{fredrik.lindsten@liu.se} \\
  \And
  Vanessa Przybylo \\
  University at Albany \\
  Albany, New York \\
  \texttt{Vanessa.Przybylo@nationalgrid.com} \\
  \And
  Kara Sulia \\
  University at Albany \\
  Albany, New York \\
  \texttt{ksulia@albany.edu} \\
  \And
  Marcus van Lier-Walqui \\
  Columbia University \\
  New York, New York \\
  \texttt{mv2525@columbia.edu} \\
  \And
  Kara D. Lamb \\
  Columbia University \\
  New York, New York \\
  \texttt{kl3231@columbia.edu}
}

\begin{document}

\maketitle

\begin{abstract}
Ice-containing clouds strongly impact climate, but they are hard to model due to ice crystal habit (i.e., shape) diversity. We use self-supervised learning (SSL) to learn latent representations of crystals from ice crystal imagery. By pre-training a vision transformer with many cloud particle images, we learn robust representations of crystal morphology, which can be used for various science-driven tasks. Our key contributions include (1) validating that our SSL approach can be used to learn meaningful representations, and (2) presenting a relevant application where we quantify ice crystal diversity with these latent representations. Our results demonstrate the power of SSL-driven representations to improve the characterization of ice crystals and subsequently constrain their role in Earth's climate system.
\end{abstract} 

\section{Introduction} \label{section:intro}
Clouds are one of the largest sources of uncertainty in climate models \citep{morrisonConfrontingChallengeModeling2020, lambPerspectivesSystematicCloud2025}. They are notoriously difficult to represent accurately in models, and ice-containing clouds are especially challenging due to highly diverse properties such as crystal morphology \citep{jarvinenAdditionalGlobalClimate2018}. Ice microphysical properties alter particle-radiation interactions and aerodynamics at the single-particle scale; and influence global radiative forcing, precipitation, and spatiotemporal distributions of clouds through a cascade of multiscale interactions \citep{shimaPredictingMorphologyIce2020, chandrakarWhatControlsCrystal2024}. Improving our understanding of clouds is crucial, since uncertainties in future cloud behavior largely drive the overall uncertainty of future climate projections \citep{bony2006well,sherwoodAssessmentEarthsClimate2020,tanModerateClimateSensitivity2025}. 

One important way to constrain ice microphysical properties is to take in situ measurements. For example, to understand the distribution of ice crystal habit (i.e., shape), millions of images of cloud particles have been taken on numerous airborne campaigns using cloud particle imagers (CPI) \citep{bakerImprovementDeterminationIce2006}. Historically, image processing techniques have been used to extract microphysically-relevant properties from CPI images \citep{mcfarquharProcessingIceCloud2017, lawsonReviewIceParticle2019}, and more recently, supervised ML has been used to improve predictions of particle properties \citep{przybyloClassificationCloudParticle2022}. However, unsupervised ML has largely been underutilized in the context of analyzing CPI data and in situ microphysical observations more broadly. 

To our knowledge, this is the first application of self-supervised learning to explore patterns of latent ice crystal representations. We pre-train a state-of-the-art vision transformer on a large CPI dataset to learn robust crystal representations that can support downstream science-oriented tasks. We also demonstrate an efficient pre-training pipeline that leverages existing pre-trained models and data curation. Model validation using a smaller, labeled test set confirms that the representations are encoding physically meaningful features. This work highlights the benefits of learning robust crystal representations and paves the way for more accurate, data-driven ice microphysical models.


\section{Data and methods} \label{section:methods}
\subsection{Dataset description}
\begin{figure}[htbp]
  \centering
  \includegraphics[width=0.65\textwidth]{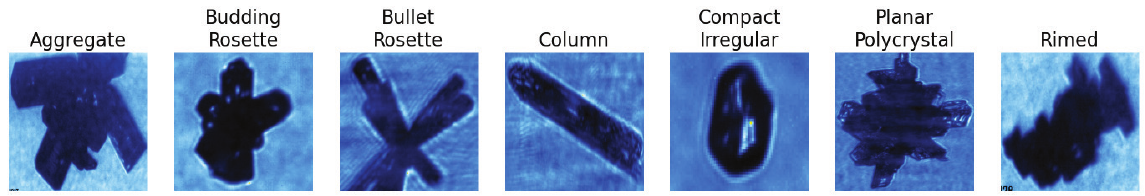}
  \caption{Examples of CPI images grouped by habit (i.e., shape).}
  \label{fig:cpi-array}
\end{figure}

The main data in this study are CPI images that come from various federally-funded airborne field campaigns. In brief, a CPI is an optical imager that takes single-channel images of cloud particles with a charge-coupled device (CCD) camera. The native CPI resolution is \SI{2.3}{\micro\meter}, but each image was resized to a resolution of \(224 \times 224\) pixels. In total, $\sim$3.2 million unlabeled CPI images from across 13 field campaigns were used as the available pre-training dataset for our model (hereafter CPI-3M). To validate learned representations, we used a smaller, hand-labeled subset of $\sim$21,000 CPI images (hereafter CPI-21K). In addition, the CPI-3M also contained habit classification labels that were predicted using a fine-tuned VGG16 convolutional neural network from \citet{przybyloClassificationCloudParticle2022}. Examples of CPI images are shown in Figure \ref{fig:cpi-array}. A subset of CPI-3M had corresponding environmental data that we used for downstream analysis (see Section \ref{section:results}). The  environmental data include measurements such as pressure, temperature, and ice water content. $\sim$524,000 CPI images had corresponding environmental measurements (hereafter CPI-ENV-500K).

\subsection{Efficient self-supervised pre-training}
\label{sec: efficient_ssl}

Self-supervised learning (SSL) is an effective approach to learn informative representations of unlabeled data \citep{ericssonSelfSupervisedRepresentationLearning2022, kolesnikovRevisitingSelfSupervisedVisual2019}. Such representations enable various downstream analyses as well as efficient predictions with minimum labeling efforts \citep{msn}. CPI images exhibit natural clusters characterized by ice habits. Hence, we consider state-of-the-art, clustering-based SSL methods from the DINO family \citep{dino, ibot}, which are trained to assign each image to a cluster such that its augmented views, obtained by applying data augmentations, are also assigned to the same cluster. \citet{dino_vmf} showed that the representations constitute a mixture model of von Mises-Fisher (vMF) distributions. We use the iBOT-vMF method \citep{dino_vmf} to pre-train our models and use the small Vision Transformer model architecture with a patch size of $16$ (more details in \ref{sec: app_ssl_details}). We evaluate the learned representations with the downstream task of classifying the CPI-21K dataset.

SSL models pre-trained on ImageNet \citep{deng2009imagenet} are publicly available. These models transfer well to ImageNet-related domains, but their performance in entirely new domains is unclear \citep{ssl_data_diversity}. First, we evaluate a CPI-3M pre-trained model and compare it with the best ImageNet pre-trained model (see Table \ref{table:ssl_methods_comparison}). We also consider the recent DINOv3 \citep{dinov3} model that is pre-trained on a larger private dataset consisting of 1.7B naturalistic images. We observe that the ImageNet pre-trained model performs well, showing that features learned from ImageNet can transfer well to CPI images. Pre-training the DINO family models on imbalanced data is a known challenge \citep{on_ppc, ssl_data_curation}. We address this limitation by following a data curation strategy proposed by \citet{ssl_data_curation}. Specifically, we curate 1.2 million images from CPI-3M through hierarchical sampling of data in the learned latent space (see \ref{sec: app_data_curation_details} for more details). This produces images that are more uniformly distributed in the latent space and hence, less imbalanced. We call this new dataset CPI-H-1M. Pre-training on this $\sim$3$\times$ smaller dataset results in an improved model, demonstrating the importance of data curation. 

The above experiments showed that data curation is important and that ImageNet pre-trained models work reasonably well on CPI data. Also, pre-training on well-curated datasets in the target domain demonstrated potential for improved performance with better evaluation results. With this motivation, we investigated if we could pre-train models for CPI data more efficiently. Specifically, we pre-train a model on the curated CPI-H-1M dataset using iBOT-vMF for only 10 epochs and initialize the model with the ImageNet pre-trained model weights. Given a well-curated dataset like CPI-H-1M, this approach is $\sim30\times$ more compute efficient than directly pre-training on the large CPI-3M dataset and resulted in the best performance based on validation of learned representations (see Table \ref{table:ssl_methods_comparison}).

\begin{table}[hbtp!]
  \caption{Comparison of self-supervised learning models on the task of classifying the CPI-21K dataset using kNN and logistic regression classifiers. We use ViT-Small model architecture for all results and report the Top-1 accuracy metric.}
  \label{table:ssl_methods_comparison}
  \centering
  \small
  \begin{tabular}{lccccc}
    \toprule
    \multirow{ 2}{*}{SSL Method} & \multirow{ 2}{*}{\makecell{Pre-training\\dataset}} &  \multirow{ 2}{*}{\makecell{Pre-training\\epochs}} & \multirow{ 2}{*}{\makecell{Weight\\initialization}} & \multicolumn{2}{c}{Top-1 Accuracy (\%)} \\
    \cmidrule{5-6}
    & & & & kNN & Logistic \\
    \midrule
    DINOv3 & LVD-1689M & 1000  & \xmark & 74.83 & 81.83 \\
    iBOT & ImageNet & 800  & \xmark & 78.33 & 82.00 \\
    iBOT-vMF & CPI-3M & 100  & \xmark & 75.05 & 81.00 \\
    iBOT-vMF & CPI-H-1M & 100  & \xmark & 77.67 & 83.17 \\
    iBOT-vMF & CPI-H-1M & 10  & \cmark & \textbf{81.56} & \textbf{84.39} \\
    \bottomrule
  \end{tabular}
\end{table}

\begin{figure}[htbp!]
  \centering
  \includegraphics[width=0.75\textwidth]{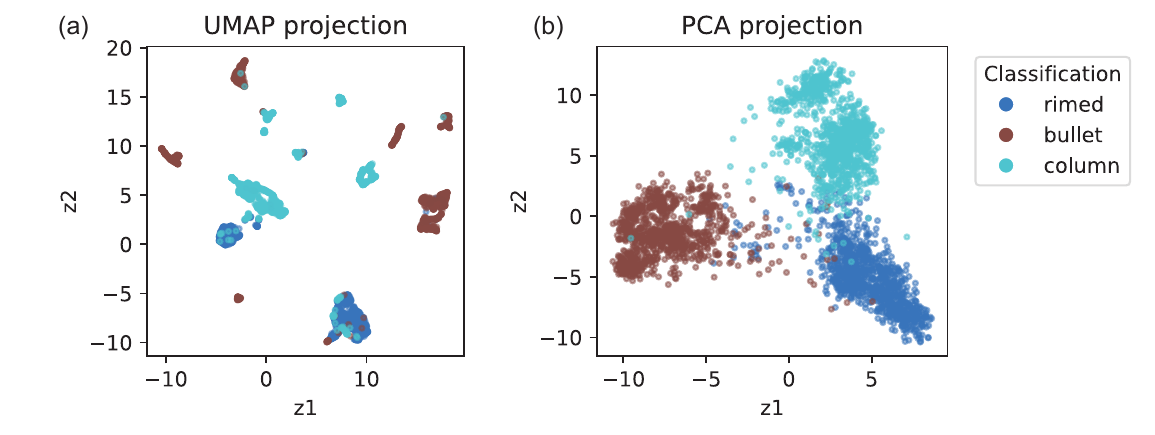}
  \caption{2D projections of the 384-dimensional latent embeddings. A subset of 3000 samples is shown here. (a) Non-linear dimensionality reduction with UMAP. (b) Linear projection with PCA.}
  \label{fig:umap-pca}
\end{figure}

\section{Results and discussion} \label{section:results}

\subsection{Quality of learned representations}\label{section:learned-reps}
To supplement the validation results in Table \ref{table:ssl_methods_comparison}, here we (1) confirm if clusters match expert habit labels, and (2) compare to a feature-extraction based baseline to demonstrate the benefit of SSL representations. For (1), we used dimensionality reduction to inspect clusters in 2D space. For clarity, we used a balanced subset of 3000 CPI-3M samples, and three of the seven classes to reduce visual clutter (see Appendix \ref{app:dim-reduce}). UMAP and PCA were used to project the 384-dimensional SSL embeddings into 2D. Figure \ref{fig:umap-pca} visualizes the projections, with points colored by labels predicted by a CNN from \citet{przybyloClassificationCloudParticle2022}. PCA reveals three distinct clusters, while UMAP forms more fragmented groupings. This aligns with the strong performance of the logistic regression on CPI-21K and suggests our model is learning approximately linearly separable morphological features without explicit guidance. For (2), we trained a baseline classifier using the CPI-21K dataset, using 13 extracted geometric features as predictors. These geometric features are derived using traditional image processing techniques, and include features such as aspect ratio, laplacian blur, and circularity, among others. Further details about these features can be found in \citet{przybyloClassificationCloudParticle2022}. The feature-based logistic regression performed with a top-1 accuracy of 65\%, which is much lower than the 84\% accuracy from our logistic regression validation using our best model (see Table \ref{table:ssl_methods_comparison}).

\subsection{Application: quantifying ice crystal diversity}\label{section:application}
\begin{figure}[htbp!]
  \centering
  \includegraphics[width=0.8\textwidth]{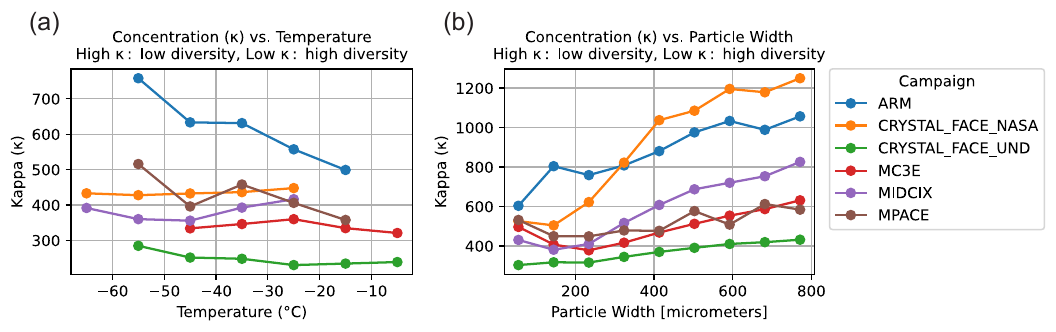}
  \caption{Crystal diversity ($\kappa$) using CPI-ENV-500K. (a) $\kappa$ as a function of air temperature and stratified by campaign. (b) $\kappa$ as a function of particle size (width) and stratified by campaign.}
  \label{fig:diversity}
\end{figure}

After validating the learned representations, we applied them to a downstream science task: quantifying ice habit diversity in real-world clouds. Existing methods to quantify habit diversity rely on pre-designated classes and assumptions about certain morphological features, whereas SSL-driven embeddings enable a purely data-driven approach without any prior assumptions. Since our representations follow vMF distributions, the most appropriate metric to characterize diversity is the $\kappa$ (i.e., ``concentration") metric \citep{sraShortNoteParameter2012} (details in Appendix \ref{app:metrics}). Using CPI-ENV-500K, we analyzed how $\kappa$ varies as a function of air temperature, particle size, and campaign. Figure \ref{fig:diversity}a generally shows increasing habit diversity with increasing temperatures, and Figure \ref{fig:diversity}b shows decreasing diversity with increasing particle size. Additionally, we see a wide spread between campaigns, highlighting the variability in crystal diversity between different cloud systems (see Appendix \ref{app:campaign}).

\begin{wrapfigure}{r}{0.6\textwidth}
  \vspace*{-0.5cm}
  \centering
  \includegraphics[width=\linewidth]{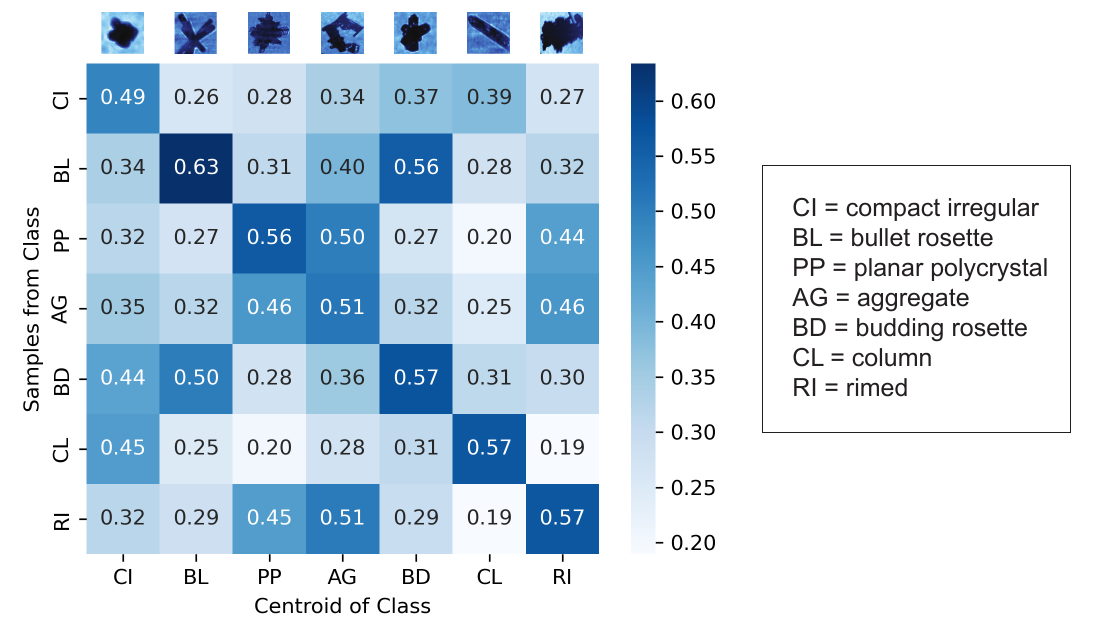}
  \caption{Cosine similarity heatmap. Intra-(diagonal) and inter-(row-wise) class similarity. Representative CPI images from each class are shown at the top.}
  \label{fig:diversity-heatmap}
  \vspace*{-0.3cm}
\end{wrapfigure}
We also quantified both intra- and inter-cluster similarity. Assuming clusters loosely follow expert labels, we computed the mean cosine similarity for each predicted class, with respect to the centroid of all other predicted classes. Figure \ref{fig:diversity-heatmap} shows the heatmap of the mean cosine similarity, where the diagonal describes the intra-cluster diversity, and the off-diagonal values describe the inter-cluster similarities. In other words, the heatmap shows us which habit classes show the most variability within that cluster, and also which clusters are most similar or dissimilar to each other.



\section{Conclusion} \label{section:conclusion}
We used the iBOT-vMF SSL vision transformer \citep{dino_vmf} to learn ice crystal representations from a large CPI dataset. Standard SSL pre-training can be computationally expensive. Through data curation and by leveraging ImageNet pre-trained weights, we outline an efficient pre-training pipeline. This resulted in robust embeddings with strong linear predictive power for downstream habit classification, validating our learned representations. These latent representations enable fully data-driven pipelines to characterize ice crystal morphology, reducing dependence on pre-defined classes and expert-driven assumptions. As a case study, we revealed systematic ice crystal diversity variations with temperature, particle size, and campaign, and quantified intra- and inter-class similarity across habit types. For future work, we will explore how learned embeddings can support anomaly detection, identifying mislabeled or rare habits, in addition to further linking microphysical properties to thermodynamic histories. This work demonstrates that SSL can effectively capture ice crystal morphology as latent representations, providing a scalable framework to reduce microphysical uncertainties and improve the representation of ice-containing clouds in climate models.

\section*{Acknowledgements}
We acknowledge funding from NSF through the Learning the Earth with Artificial Intelligence and Physics (LEAP) Science and Technology Center (STC) (Award \#2019625). This research was also financially supported by the Wallenberg AI, Autonomous Systems and Software Program (WASP), and the Excellence Center at Linköping--Lund in Information Technology (ELLIIT). The collaboration was initiated during the ELLIIT Focus Period on Machine Learning for Climate Science in Linköping, 2024, and we thank ELLIIT for the support during the program. Computations were enabled by the Berzelius resource at the National Supercomputer Centre, provided by the Knut and Alice Wallenberg Foundation. 

\bibliographystyle{unsrtnat}
\bibliography{references}

\newpage
\appendix
\section{Appendix}

\subsection{Dimensionality reduction} \label{app:dim-reduce}
\begin{figure}[htbp!]
    \centering
    \includegraphics[width=\linewidth]{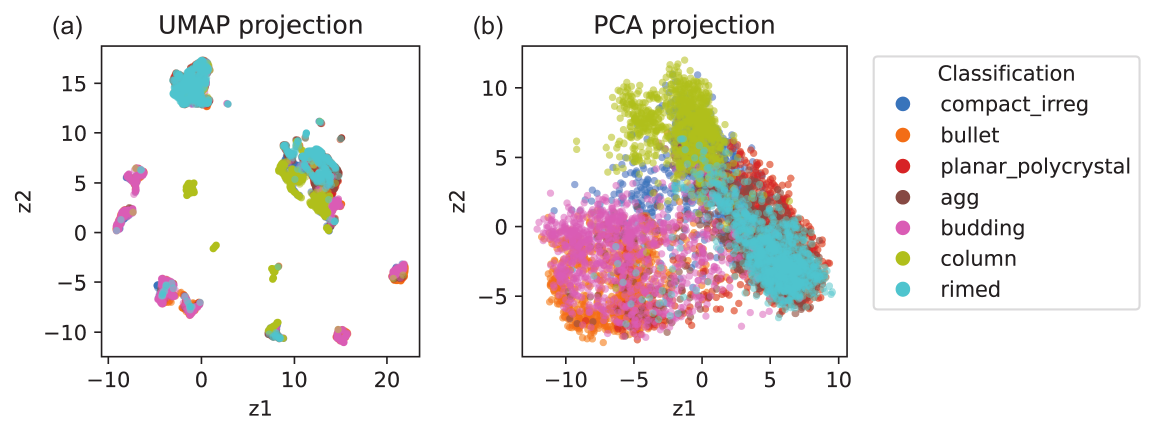}
    \caption{2D projections of the 384-dimensional latent embeddings. 7000 samples (1000 samples per class) are shown here. (a) Non-linear dimensionality reduction with UMAP. (b) Linear projection with PCA.}
    \label{fig:umpa-pca-2d-allclass}
\end{figure}

Figure \ref{fig:umpa-pca-2d-allclass} shows the 2D UMAP and PCA projections including all classes, as referenced in Section \ref{section:learned-reps}. 1000 samples per class are shown here, analogous to Figure \ref{fig:umap-pca}. Note the high degree of overlap in 2D space, which is not surprising given that 384-dimensional embeddings are being reduced to 2D. Three distinct classes were chosen out of the seven classes shown here for the main text to reduce visual clutter and for the sake of demonstration.

\subsection{Campaign details} \label{app:campaign}
\begin{figure}[htbp!]
    \centering
    \includegraphics[width=\linewidth]{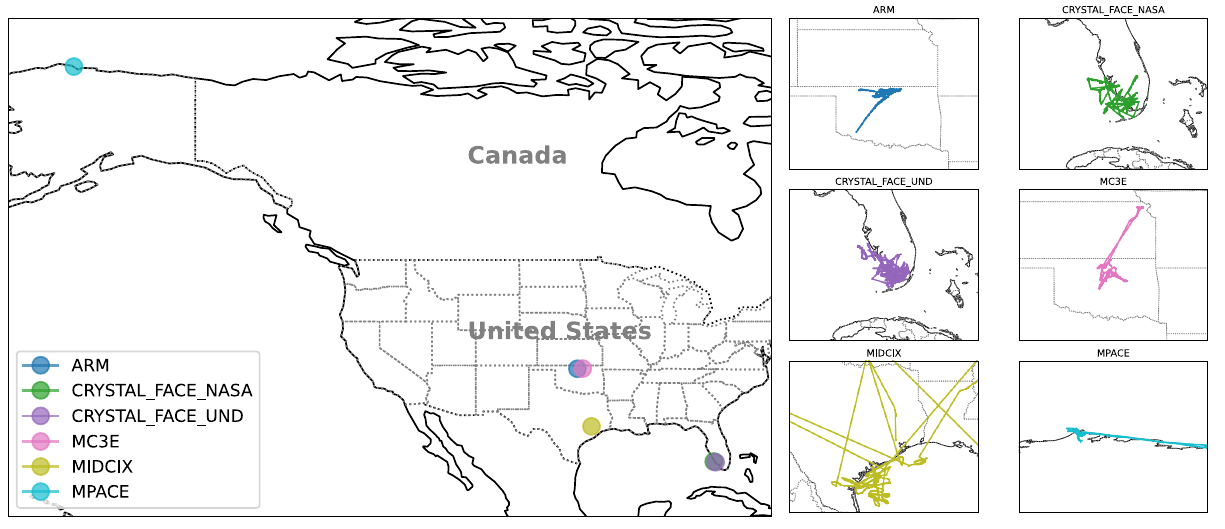}
    \caption{Maps showing the general locations (left) of the different field campaigns mentioned in Section \ref{section:application} as well as the respective zoomed-in views (right) of the flight tracks.}
    \label{fig:flight-tracks}
\end{figure}

\begin{figure}[htbp!]
    \centering
    \includegraphics[width=\linewidth]{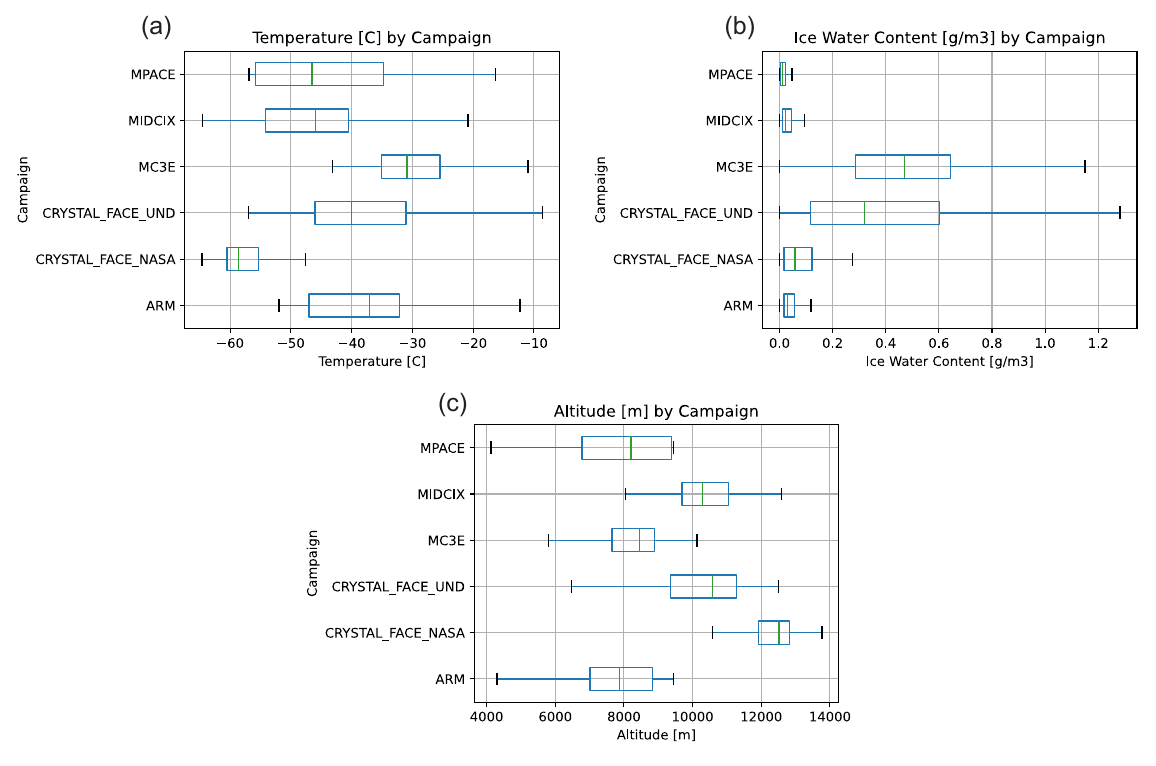}
    \caption{Distributions of (a) air temperature, (b) ice water content, and (c) altitude for the different field campaigns represented as box-and-whisker plots.}
    \label{fig:campaign-stats}
\end{figure}

Additional details regarding the campaigns are described here. Figure \ref{fig:flight-tracks} shows a map indicating where the various campaigns were conducted, as well as displaying the flight tracks of the individual campaigns in more detail. Figure \ref{fig:campaign-stats} shows the variability in conditions (e.g., temperature, ice water content, and altitude) between the different field campaigns. As mentioned in Section \ref{section:application}, we observed a wide inter-campaign range of crystal diversity. The large differences in environmental conditions for the different campaigns are highlighted here. 

\subsection{Metrics details} \label{app:metrics}

For a random $p$-dimensional unit vector $\boldsymbol{x_i}$, the von Mises-Fisher (vMF) probability distribution is given by $f(\boldsymbol{x_i}; \boldsymbol{\mu}, \kappa) = C_p(\kappa) \exp(\kappa \boldsymbol{\mu}^T \boldsymbol{x_i})$, 
where $\boldsymbol{\mu}$ is a mean vector with $\|\boldsymbol{\mu}\| = 1$, $\kappa$ is a scalar concentration parameter that measures isotropic precision, and $C_p(\kappa)$ is a normalizing constant. A higher value of the parameter $\kappa$ denotes a higher concentration of samples around the mean vector $\boldsymbol{\mu}$ and lower variance or diversity. On the other hand, a lower value of the parameter $\kappa$ denotes a lower concentration and consequently a higher variance or diversity.
We used the following equations to estimate $\kappa$ for a set of embedding vectors \citep{sraShortNoteParameter2012}:

\begin{center}
\begin{minipage}{0.25\textwidth}
\begin{equation*}
\hat{\kappa} = \frac{\bar{R}\left( p - \bar{R}^2 \right)}{1 - \bar{R}^2}
\end{equation*}
\end{minipage}
\hspace{0.02\textwidth}
\begin{minipage}{0.25\textwidth}
\begin{equation*}
\bar{R} = \frac{\left\| \sum_{i=1}^N x_i \right\|}{N}
\end{equation*}
\end{minipage}
\end{center}

where $x_i \in \mathbb{R}^p$ is the $i$-th normalized embedding vector, $N$ is the number of samples, and $p$ is the embedding dimensionality.

We computed cosine similarity between two embedding vectors $x_i, x_j \in \mathbb{R}^p$ as:  

\begin{center}
\begin{minipage}{0.25\textwidth}
\begin{equation*}
\mathrm{cosine\_similarity}(x_i, x_j) = \frac{x_i \cdot x_j}{\|x_i\| \, \|x_j\|}
\end{equation*}
\end{minipage}
\end{center}

\subsection{Datasets}
\label{sec: app_datasets_details}

The CPI-3M dataset is a raw uncurated dataset obtained by combining images from all airborne field campaigns. The CPI-H-1M is curated in an unsupervised manner from the CPI-3M dataset, as detailed in the Section \ref{sec: efficient_ssl} and Appendix \ref{sec: app_data_curation_details}. A subset of data from CPI-3M dataset contained environmental data and this subset is denoted as CPI-ENV-500K. CPI-21K is a smaller dataset where the ice habit classes are hand-labeled. This dataset is further split into training and test splits consisting of 19,737 and 1,800 images respectively. The test split is balanced and contains equal number of images per class, that is, 200 images in each of the nine classes.

\begin{table}[hbtp!]
  \caption{CPI Image Datasets. Pseudo-labeled means labels were predicted using a supervised CNN described in \citet{przybyloClassificationCloudParticle2022}.}
  \label{table:cpi_datasets}
  \centering
  \small
  \begin{tabular}{lccccc}
    \toprule
    Dataset name & \# images &  Labels & Label type & Environmental data & Data curation \\
    \midrule
    CPI-3M & 3,200,351 & \cmark & Pseudo-labeled using a CNN & \xmark & \xmark \\
    CPI-H-1M & 1,200,000 & \cmark & Pseudo-labeled using a CNN & \xmark & \cmark \\
    CPI-ENV-500K & 524,000 & \cmark & Pseudo-labeled using a CNN & \cmark & \xmark \\
    CPI-21K & 21,537 & \cmark & Hand-labeled & \xmark & \xmark \\
    \bottomrule
  \end{tabular}
\end{table}

\subsection{Implementation details}
\label{sec: app_implementation_details}

\subsubsection{Self-supervised Pre-training}
\label{sec: app_ssl_details}

\paragraph{Method overview:} We use a clustering-based self-supervised pre-training method from the DINO family, known as iBOT \citep{ibot}. This uses a teacher-student self-distillation setup where both the teacher and the student have the same model architecture and are initialized with random weights. Given images $\vx$ from an unlabeled image dataset, we obtain randomly augmented views of the image, $\vx_s = A_s(\vx)$ and $\vx_t = A_t(\vx)$. Here, $A_s$ and $A_t$ are random augmentations specific to the student and teacher models. The student and teacher networks consist of a backbone model (typically a Vision Transformer) and a prediction head. The backbone model is used for different downstream tasks but the prediction head (typically an MLP) is only used during pre-training and then discarded. The student and teacher models output probability distributions over $K$ pseudo-classes or clusters. The student model is trained to match the cluster assignment of the teacher, using a cross-entropy loss between the student and teacher outputs. The student model weights are updated using the loss gradients and the teacher model is updated using an exponential moving average of the student model weights. For a more detailed description and motivation behind this training methodology, we refer the reader to the original papers of DINO \citep{dino}, iBOT \citep{ibot} and DINO-vMF \citep{dino_vmf}.

\paragraph{Implementation:} We use the public codebase of iBOT \footnote{https://github.com/bytedance/ibot/} and use the vMF normalized formulation proposed in \citet{dino_vmf}. We modified the image augmentations used during the pre-training based on their suitability to CPI images. Since, CPI images are monochromatic, we remove the jitter to the image saturation and hue. In addition to the random horizontal flip, we added a random vertical flip with a probability of $0.5$, as the crystals can be freely rotated in space. In the existing random resized crop augmentation, we reduced the change in aspect ratio of the crop by setting the new aspect ratio to be in the range of $(0.9, 1.1)$. The ice crystals contain spikes of varying thickness which is an important distinguishing feature of the crystal and we want this information to be preserved in the learned representations. We use this modified set of augmentations for all the pre-training experiments that we conducted. Other hyperparameter settings for both the standard pre-training setup and the shorter and more efficient pre-training setup are provided in Table \ref{table:app_hyperparams_ibot_vmf}. In the efficient pre-training setup, we initialize the model with the weights from a model pre-trained on ImageNet-1K dataset using the iBOT method \citep{ibot}. 

\begin{table}[h]
  \caption{Hyperparameter settings for iBOT-vMF pre-training using the standard setup and the proposed efficient setup using weights initialized from an ImageNet pre-training.}
  \label{table:app_hyperparams_ibot_vmf}
  \centering
  \begin{tabular}{lll}
    \toprule
    Hyperparameter & Standard iBOT-vMF & Efficient iBOT-vMF \\
    \midrule
    training epochs & $100$ & $10$ \\
    batch size & $1024$ & $1024$ \\
    learning rate & $4\sce{-4}$ & $3\sce{-4}$ \\
    warmup epochs & $10$ & $8$ \\
    freeze last layer epochs & $1$ & $1$ \\
    min. learning rate & $1\sce{-6}$ & $1\sce{-6}$ \\
    weight decay & $0.04 \rightarrow 0.4$ & $0.04 \rightarrow 0.1$ \\
    stochastic depth & $0.1$ & $0.1$ \\
    gradient clip & $1.0$ & $1.0$ \\
    optimizer & adamw & adamw \\
    shared head & \cmark & \cmark \\
    fp16 & \cmark & \cmark \\
    \midrule
    momentum & $0.996 \rightarrow 1.0$ & $0.996 \rightarrow 1.0$ \\
    global crops & $2$ & $2$ \\
    global crops scale & $[0.32, 1.0]$ & $[0.32, 1.0]$ \\
    local crops & $10$ & $10$ \\
    local crops scale & $[0.1, 0.32]$ & $[0.1, 0.32]$ \\
    \midrule
    head mlp layers & $3$ & $3$ \\
    head hidden dim. & $1024$ & $1024$ \\
    head bottleneck dim. & $64$ & $64$ \\
    norm last layer & \xmark & \xmark \\
    num. prototypes & $2048$ & $2048$ \\
    vmf normalization & \cmark & \cmark \\
    centering & probability & probability \\
    \midrule
    teacher temp. & $0.04 \rightarrow 0.07$ & $0.04$ \\
    temp. warmup epochs & $30$ & $-$ \\
    student temp. & $0.1$ & $0.1$ \\
    \midrule
    pred. ratio & $[0.0, 0.3]$ & $[0.0, 0.3]$ \\
    pred. ratio variance & $[0.0, 0.2]$ & $[0.0, 0.2]$ \\
    pred. shape & block & block \\
    \bottomrule
  \end{tabular}
\end{table}

\subsubsection{Data curation}
\label{sec: app_data_curation_details}

In this section, we provide additional details on how we curate the CPI-H-1M dataset from the larger CPI-3M dataset. Firstly, we run a hierarchical KMeans algorithm on the latent representations using a hierarchy as follows: 3.2M images $\rightarrow$ 50K clusters $\rightarrow$ 5K clusters $\rightarrow$ 1K clusters $\rightarrow$ 200 clusters. We use the latent representations obtained from the ViT model pre-trained using iBOT-vMF on CPI-3M dataset. If an existing pre-trained model (such as those trained on ImageNet) would perform reasonably well on the target dataset, then one could also consider those latent representations for this step. Then, we use hierarchical sampling where we compute the number of samples per sub-tree, starting from the coarsest level in the hierarchy. As demonstrated in \citet{ssl_data_curation}, this produces uniform distribution of samples across different levels in the hierarchy. We used the code available in their public repository \footnote{https://github.com/facebookresearch/ssl-data-curation}.

\end{document}

%% file: math_commands.tex
\usepackage{amsmath,amsfonts,bm}
\usepackage{xspace}
\usepackage{pifont}

\def\eqref#1{equation~\ref{#1}}

\def\1{\bm{1}}

\def\vx{{\bm{x}}}

\DeclareMathAlphabet{\mathsfit}{\encodingdefault}{\sfdefault}{m}{sl}
\SetMathAlphabet{\mathsfit}{bold}{\encodingdefault}{\sfdefault}{bx}{n}

\newcommand{\cmark}{\ding{51}}%
\newcommand{\xmark}{\ding{55}}%

\newcommand{\sce}{\mathrm{e}}